\documentclass{aastex631}
\usepackage{natbib}

\shorttitle{LSST Survey Strategies and Brown Dwarf Parallaxes}
\shortauthors{Gizis et al.}

\graphicspath{{./}}

\begin{document}
\title{LSST Survey Strategies and Brown Dwarf Parallaxes}

\author[0000-0002-8916-1972]{John E. Gizis}
\affil{Department of Physics and Astronomy, University of Delaware, Newark, DE 19716, USA}

\author[0000-0003-2874-6464]{Peter Yoachim}
\affil{Department of Astronomy, University of Washington, 3910 15th Ave. NE, Seattle, WA 98195, USA}

\author[0000-0001-5916-0031]{R. Lynne Jones}
\affil{Aerotek and Rubin Observatory, Tucson, AZ, USA}

\author[0000-0002-7950-0651]{Dylan Hilligoss} 
\affil{Department of Physics and Astronomy, University of Delaware, Newark, DE 19716, USA}

\author[0000-0003-0906-1195]{Jinbiao Ji}
\affil{Department of Physics and Astronomy, University of Delaware, Newark, DE 19716, USA}

\begin{abstract}
The Vera C. Rubin Observatory's Legacy Survey of Space and Time (LSST) has the potential to measure parallaxes for thousands of nearby ultracool dwarfs, enabling improved measurements of the brown dwarf luminosity function. We develop a simple model to estimate the number of L dwarfs and T dwarfs with parallaxes with signal-to-noise ratio greater than ten in the baseline LSST survey. High quality astrometry imposes scheduling constraints. We assess different possible observing strategies using quantitative metrics and make recommendations as part of the LSST community input process. We find that the new substellar parallax sample will represent a nearly order-of-magnitude increase on existing samples, with $\sim50-100$ objects per spectral type bin for late-L to mid-T dwarfs. The sample size is robust ($\pm 5\%$ or less) against most survey strategy changes under consideration, although we do identify areas of tension with other uses of twilight time that could have larger impact.
\end{abstract}

\keywords{Brown dwarfs (185), Sky surveys (1464), L dwarfs (894), T dwarfs (1679), Parallax (1197)}

\section{Introduction}

``Mapping the Milky Way" is one of the four primary science goals 
for the Vera C. Rubin Observatory's Legacy Survey of Space and Time (LSST) \citep{2019ApJ...873..111I}. The study of the Galaxy and the Solar Neighborhood has been revolutionized by the parallaxes, proper motions, and photometry produced by the Gaia mission \citep{2016A&A...595A...1G}. Typical parallax uncertainties are 0.07 mas at $G=17$ mag and 1.0 mas at $G=21$ mag in Gaia Early Data Release 3 \citep{2021A&A...649A...1G}.  LSST individual visits reach limiting magnitudes of $r \approx 24.5$ (5 sigma), well beyond the Gaia faint limit ($G \approx 21$ mag).    
\citet{1983AJ.....88.1489M} first demonstrated that ground-based telescopes with Charge Coupled Devices (CCDs) are capable of mas precision parallaxes and the Pan-STARRS survey has demonstrated parallaxes from ground-based sky survey data \citep{2020ApJS..251....6M,2020RNAAS...4...54D}.
The ten-year, six-filter ($ugrizy$) LSST survey thus offers the opportunity to measure astrometry for many stars too faint for Gaia. 
The astrometric specifications in the LSST System Science Requirements Document (SRD)\footnote{The LSST Science Requirements Document is available at \url{https://ls.st/srd}} were set by the overall aim of achieving proper motion accuracy of 0.2 mas/year and parallax accuracy of 1.0 mas for high signal-to-noise sources \citep{2019ApJ...873..111I}. It is expected these aims can be met in the main (Wide-Fast-Deep, hereafter WFD) survey with 825 visits over 18,000 square degrees and the specified per visit 10 mas astrometric uncertainty. The exact scheduling of the survey is flexible under the SRD but it recognizes that measuring parallaxes places ``strong constraints on the distribution of visits in time." These include the need for a long baseline in time rather than clustering observations at a few epochs, observing near the meridian early and late in the night to sample high parallax factors, and minimizing differential color refraction (DCR) by observing at low airmass and avoiding correlations between the parallax and atmospheric refraction angles.

The Vera C. Rubin Observatory is seeking to optimize the observing strategy through a community process described by \citet{2022ApJS..258....1B}. The operations simulator  \citep[\texttt{OpSim}]{2014SPIE.9150E..15D} is used to model each visit of the ten year LSST survey. Following a set of recommendations by its Science Advisory Committee (SAC), the Rubin Observatory created 173 \texttt{OpSim} simulated surveys, scheduled using the Feature Based Scheduler \citep{2019AJ....157..151N}, considering a range of possible strategies \citep{jones_r_lynne_2020_4048838}. The LSST Metric Analysis Framework \citep[\texttt{MAF}]{2014SPIE.9149E..0BJ} provides a set of Python-based tools to analyze these simulations. The standard metrics include assessments of the proper motion and parallax accuracy for bright sources seen in all filters. The Survey Cadence Optimization Committee (SCOC) called for science-driven notes addressing specific questions with quantitative metrics based on these simulations. The 43 notes received in 2021 are the basis for the papers in this Focus Issue of the Astrophysical Journal Supplement. The LSST Stars, Milky Way and Local Volume (SMWLV) Science Collaboration contributed a number of these notes. Here, we examine the specific science case of measuring the brown dwarf luminosity function with LSST parallaxes.\footnote{Other science cases for very-low-mass stars and brown dwarfs, such as star counts, rotation, eclipsing binaries, and flares, are discussed in the LSST Science Book \citep{2009arXiv0912.0201L}. We believe most of these are not as sensitive to observing strategies as parallaxes, or are covered by general stellar variability metrics.} Unlike many SMWLV science cases which emphasized the Galactic Plane or Bulge, all directions on the sky are valuable because the targets lie in the immediate solar neighborhood ($<50-100$ pc). 

It is now estimated that the star formation process creates one brown dwarf below the hydrogen-burning limit ($\sim0.078$ solar masses, \citep{2019ApJ...879...94F}) for every six stars above the hydrogen-burning limit \citep{2012ApJ...753..156K}. 
Old, very-low-mass hydrogen-burning stars and brown dwarfs can have effectively indistinguishable luminosities and effective temperatures, so the more general term ``ultracool dwarf" for spectral types M7 and cooler is often adopted. Measurements of the luminosity function --- the number of ultracool dwarfs per cubic parsec (space density) as a function of their absolute magnitude --- is a key tool in investigating their physics. For simplicity, we will also call measurements of the space density as a function of spectral type the ``luminosity function." The luminosity function depends on the mass function (itself a result of star formation at the extreme low-mass end), Galactic star formation and dynamical history, and stellar (or substellar) evolution that in turn depends on both interior and atmosphere physics.  

A luminosity function that is selected by observed parallaxes should be less biased than one selected on colors, proper motions, or other criteria. The Gaia Catalog of Nearby Stars
\citep[GCNS]{2021A&A...649A...6G} is a parallax-selected catalog that includes over 330,000 stars, representing an estimated 92\% of stars within 100 pc with spectral type M9 or earlier. 
Gaia has also been used to discover many previously unknown L dwarfs by their parallaxes \citep{2018A&A...619L...8R}. 
However, for the cooler spectral types L and T, Gaia becomes incomplete at closer limits: GCNS estimates its L8 dwarf sample is only complete to 10 pc. \citet{2021ApJS..253....7K} and \citet{2021AJ....161...42B} have developed and analyzed parallax-based catalogs of nearby ultracool dwarfs, with distance limits of 15-25 pc. 
These are largely based on Gaia parallaxes \citep{2018A&A...616A...1G} for early-L dwarfs, targeted ground-based parallaxes of known L and T dwarfs \citep[such as][]{2020AJ....159..257B}, and a targeted Spitzer mid-infrared parallax program for late-T and Y dwarfs \citep{2013AAS...22115830D,2019ApJS..240...19K,2021ApJS..253....7K}. 
Parallaxes can also be measured in the mid-infrared for ultracool dwarfs within 20 pc directly from the WISE survey \citep{2018ApJ...862..173T}.  

Analyses of the current luminosity function demonstrates its potential as a probe of astrophysics but point to the need for larger samples. A dip around spectral type L3 may be linked to the hydrogen-burning limit \citep{2019ApJ...883..205B,2021A&A...649A...6G}. At cooler temperatures, the L/T transition is a key problem. \citet{2008ApJ...689.1327S} predict a pileup at 1300K due to atmospheric cloud clearing changing the cooling rate -- a prediction that is confirmed by \citet{2021ApJS..253....7K}  using very wide bins in effective temperature but complicated by \citet{2021AJ....161...42B}, who instead emphasize a narrow gap at 1300K and pileups elsewhere. In any case, it is clear the observed luminosity function supports the idea that atmospheric clouds and clearing play a key role in brown dwarf evolution. Further analysis is hampered by small statistical samples --- some $\sim10-20$ objects per spectral type bin for late L and early T dwarfs --- that in turn depend on selection by photometry and/or proper motion of previous sky surveys, which will be addressed by LSST’s larger samples. 

In this paper, we present an analysis of LSST and the brown dwarf luminosity function. We develop a simplified \texttt{MAF} model to estimate the number of ultracool dwarfs as a function of spectral type that will have high signal-to-noise parallaxes in the baseline LSST survey. We calculate metrics to evaluate the relative performance of different survey strategies for this science case, and make recommendations for the survey strategy. These results demonstrate that the Rubin Observatory LSST should produce a significant increase to current ultracool dwarf parallax samples.  

\section{A model of LSST brown dwarf astrometry and the baseline case}

The LSST Version 1.7 baseline case (\texttt{baseline\_nexp2\_v1.7\_10yrs}) consists of a southern hemisphere WFD survey footprint including the outer galactic plane with additional ``mini-surveys" of the inner (crowded) Galactic Plane and Bulge, South Celestial Cap, and North Ecliptic Spur which receive fewer visits. Five Deep Drilling Fields (DDFs) fields, all at high galactic latitude, and chosen for their complementary deep extragalactic datasets, are visited much more often. All visits consist of a pair of 15 s exposures taken in the same filter. We aim to calculate the number of brown dwarfs with parallax signal-to-noise of ten or greater in this LSST baseline survey. Lower signal-to-noise parallaxes will be valuable but their use is subject to significant systematic corrections \citep{1973PASP...85..573L,2015PASP..127..994B}. 

For each spectral type, we adopt the mean Pan-STARRS (PS1) $i$, $z$, and $y$ absolute magnitudes calculated by \citet{2018ApJS..234....1B} --- neglecting any systematic difference between the PS1 and the LSST photometric systems and the dispersion of absolute magnitudes and colors at a given spectral type. We also assume all sources are single and neglect the effect of unresolved binaries. 

\begin{figure}[ht!]
\plotone{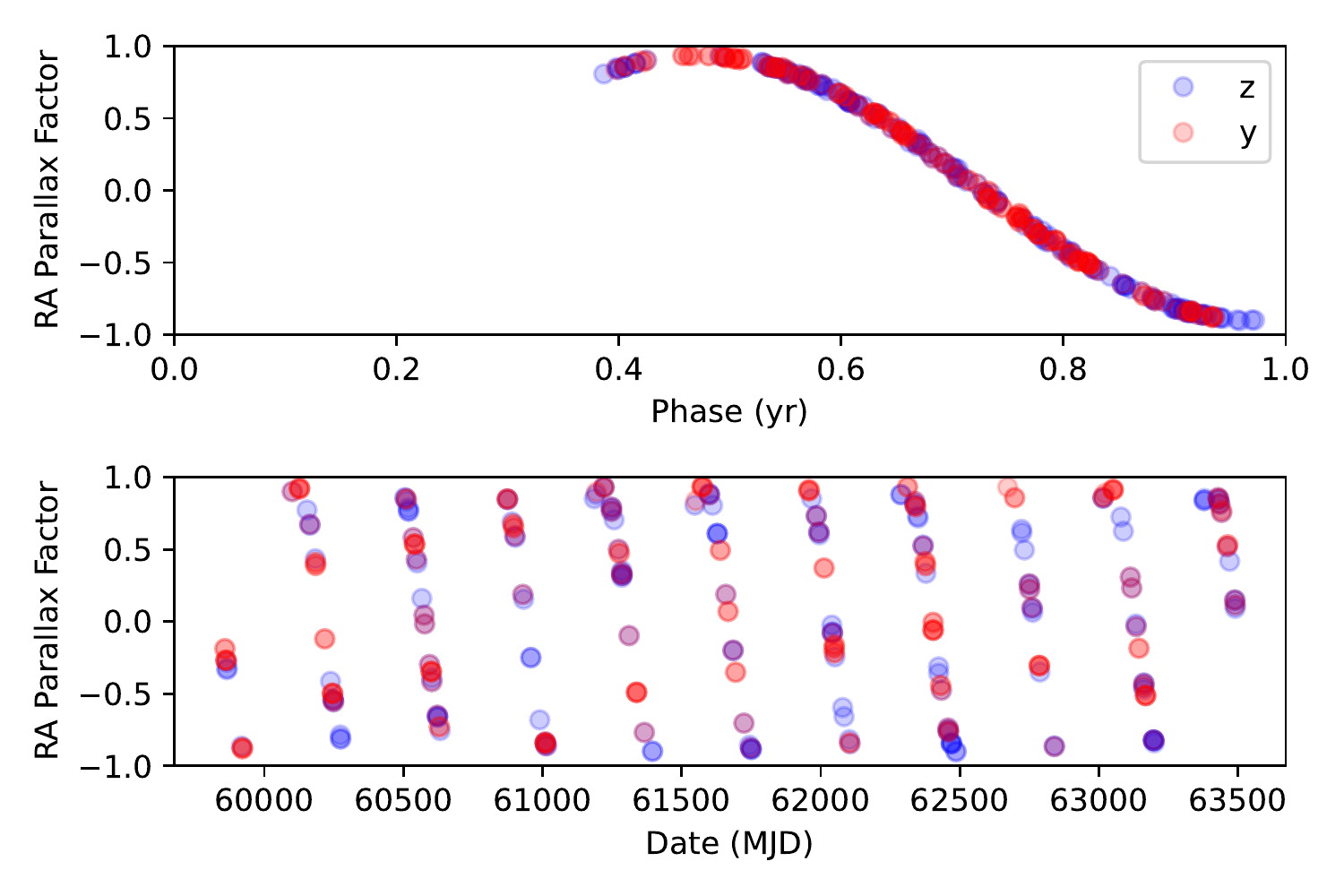}
\caption{The parallax factors sampled in $z$ and $y$ bands at an representative location ($0^h$, $-20^d$) in the WFD survey. This point receives 345 visits in $z$ and $y$. The shift in position at each epoch for each star will be its parallax (one over its distance in parsecs) times the parallax factor. Observations near $+1$ and $-1$ are necessary to fit the parallax. Top: The observations phased by year. Bottom: The observations as a function of time.   \label{fig:parallax-example}}
\end{figure}

\begin{figure}
\plotone{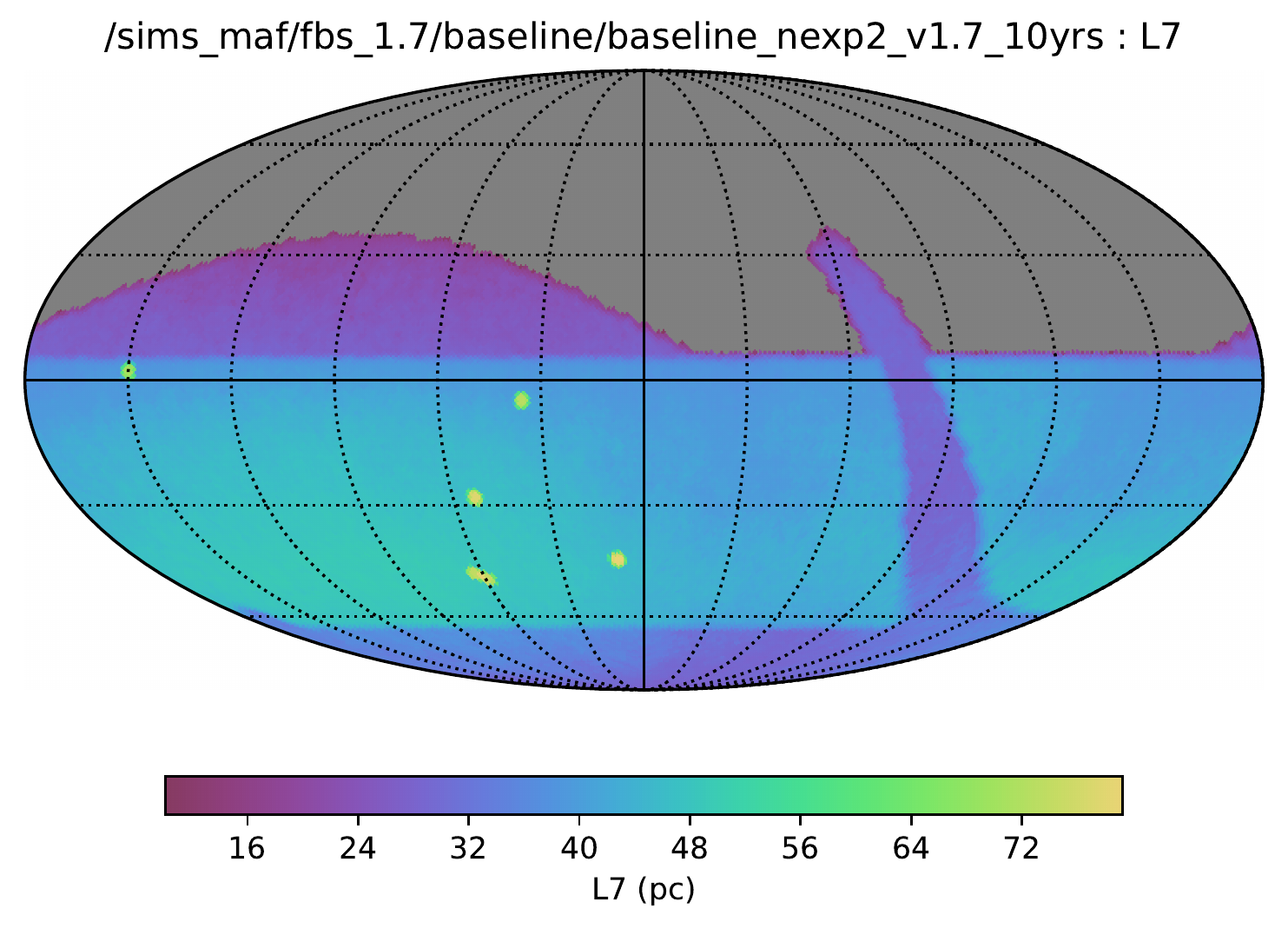}
\caption{The sky map of the limiting distance for parallax signal-to-noise ratio of 10 for L7 dwarfs produced by \texttt{MAF} for the V1.7 baseline. Most of the WFD area reaches 40-50pc for L7 dwarfs.  The inner galactic plane, south celestial gap, and north ecliptic spur regions are less deep, and the five extragalactic deep drilling fields are deeper. Spectral types M6 through T9 were modeled. The complete figure set (24 images) is available in the online journal. \label{fig-baseline-sky} }
\end{figure}

\begin{figure}
\plotone{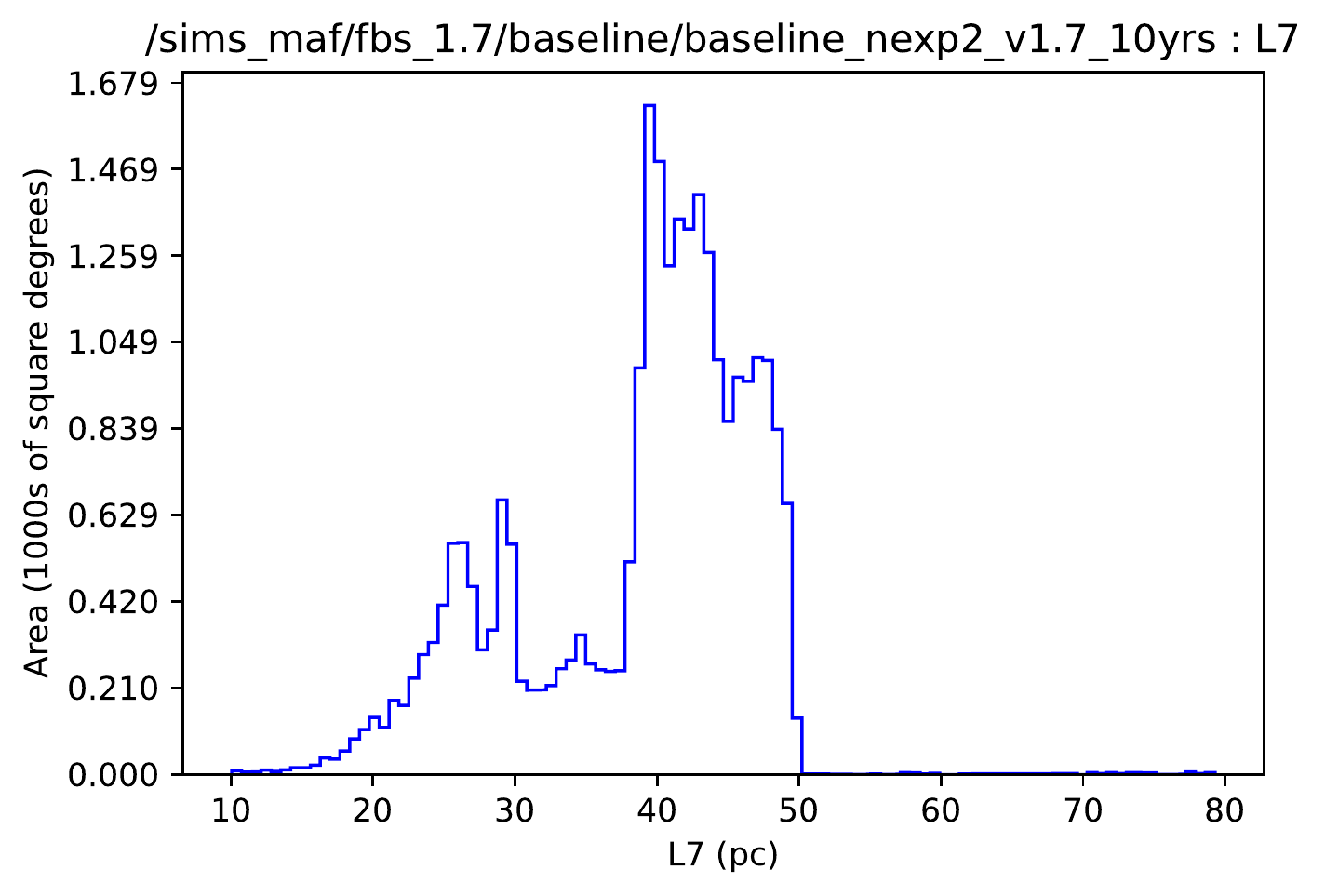}
\caption{A histogram showing the area on the sky surveyed as a function of the limiting distance for L7 dwarfs produced by \texttt{MAF} for the V1.7 baseline. Spectral types M6 through T9 were modeled. The complete figure set (24 images) is available in the online journal.\label{fig-baseline-hist}} 
\end{figure}

\begin{figure}
\plotone{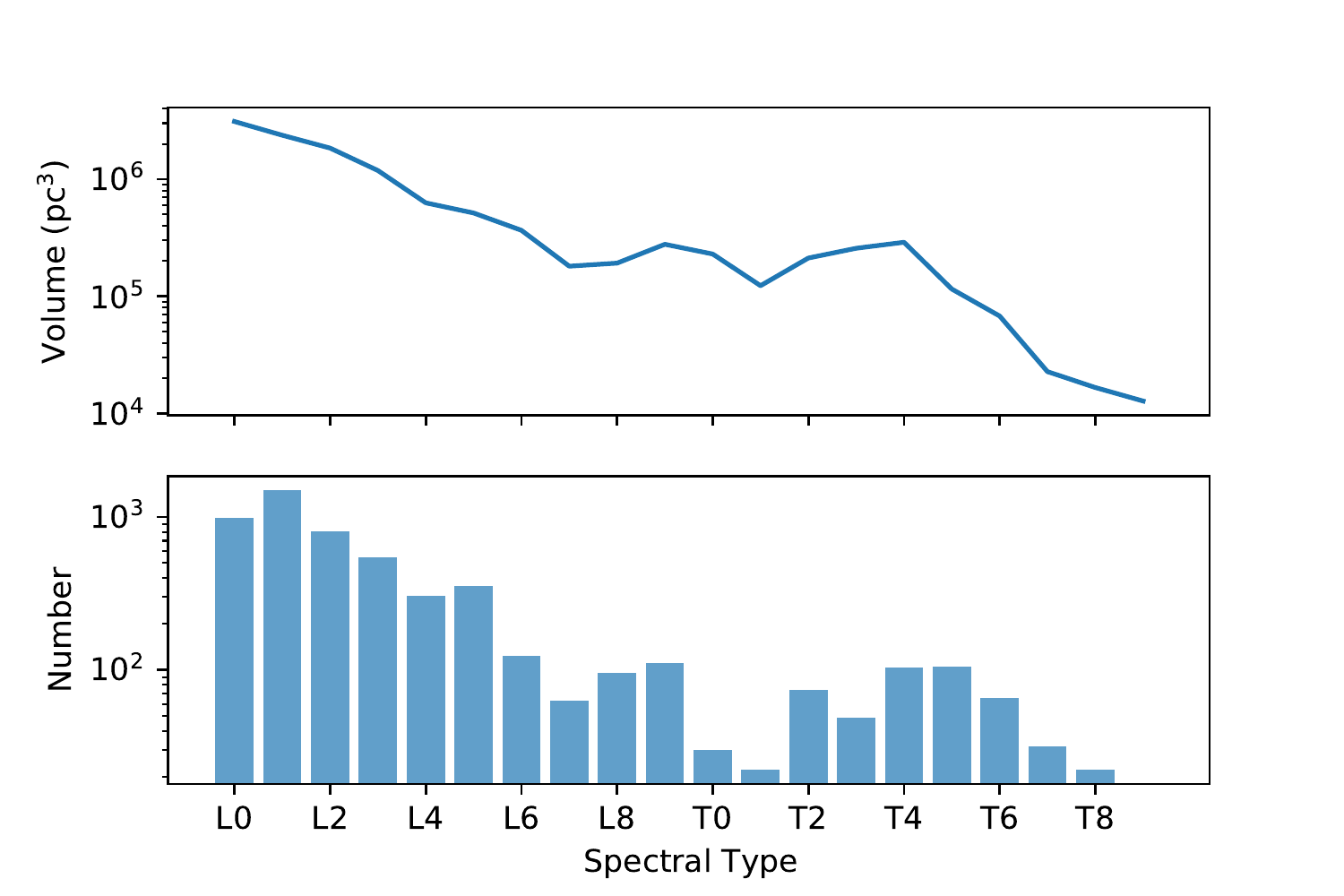}
\caption{The surveyed volume in cubic parsecs and corresponding number of L and T dwarfs with parallax SNR$\ge 10$ using the \citep{2021AJ....161...42B} space densities for the V1/7 baseline case. \label{fig-numbers} }
\end{figure}

We illustrate our procedure\footnote{The Python notebooks with all code used for this paper are available at \url{https://github.com/jgizis/LSST-BD-Cadence}. The \texttt{MAF} calculations were run on the the NOIRLab Astro Data Lab science platform at \url{https://datalab.noirlab.edu/}} with the LSST V1.7 baseline cadence  and L7 dwarfs with absolute magnitude $M_i = 20.09$ mag, $M_z = 18.18$ mag, and $M_y = 17.13$ mag.\footnote{L7 includes L7.0 and L7.5 dwarfs. The same binning is true for all spectral types discussed in this paper.} We divide the sky into healpix (Nside$=64$). For every visit, 
\texttt{MAF} has the filter, the time, the parallax factors in right ascension (RA) and declination, the seeing full width at half max (FWHM), and the photometric signal-to-noise ratio (SNR) for any given apparent magnitude. The parallax factor for a representative point on the sky in the V1.7 baseline is shown in Figure~\ref{fig:parallax-example}. The astrometric uncertainty in each coordinate for each visit is assumed to be the seeing FWHM divided by the photometric SNR (i.e., the centroiding uncertainty due to photon shot noise) added in quadrature to the atmospheric/instrumental uncertainty of 10 mas. Thus, we have all the simulated data needed to compute the parallax uncertainty for a source located at each healpix, if we know its apparent magnitudes.\footnote{The \texttt{MAF} analytic calculation of parallax uncertainty uses the individual uncertainties weighted by parallax factor, and neglects the degeneracy in fitting the position, proper motions and differential color refraction, but more detailed simulations have shown that this is a small effect for hundreds for visits (D. Monet, priv. comm.).} We proceed by computing the parallax, the parallax uncertainty, and the parallax signal-to-noise ratio for L7 dwarfs placed at different distances (10pc, 30pc, 50pc, etc.) at that healpix for that simulation. We interpolate to determine the limiting distance at which the parallax SNR is 10. We repeat the calculation for every healpix, and then sum the area of each healpix times the distance cubed to calculate the total volume probed. The distance limit projected on the sky is shown in Figure~\ref{fig-baseline-sky} and the histogram of area as a function of distance limit is shown in Figure~\ref{fig-baseline-hist}. For the V1.7 baseline cadence, the volume for L7 dwarfs is $1.8 \times 10^5$ {\rm pc}$^3$. \footnote{We have also simulated the V1.7 baseline for unresolved, equal-luminosity L7 binaries. In this case, the observed volume is 1.7 times larger, a significant increase but less than the $2^{3/2} = 2.8$ increase expected if only magnitude-limited. A detailed accounting of Malmquist biases due to unresolved binaries will require knowledge of the actual LSST performance and a model of the secondary's magnitude differences.}
 Using \citet{2021ApJS..253....7K}'s space density of 0.00060 $\rm{pc}^{-3}$ (based on 19 objects), we estimate LSST would be able to measure $\sim110$ L7 dwarfs. As illustrated in Figures~\ref{fig-baseline-sky} and~\ref{fig-baseline-hist}, the main contribution to the sample is the WFD survey. The five deep drilling fields reach greater distances due to the much larger number of observations, but they have such a small area that they make only a small contribution to the total volume. The mini-surveys of the inner Galactic plane, south celestial pole, and north ecliptic spur have fewer visits and consequently reach to smaller distances, as evident in Figure~\ref{fig-baseline-sky}. For the V1.7 baseline survey, we repeat this calculation for spectral types M6-T9 and the results are shown in the online versions of Figures~\ref{fig-baseline-sky} and~\ref{fig-baseline-hist}. The M dwarf limiting distances are well beyond 100pc, benefiting both from their brighter absolute magnitudes in $z$ and $y$ and adding useful visits in $r$ and $i$, but this also depends on whether LSST can really achieve parallax uncertainties $<1$ mas. The L and T dwarf results are listed in Table~\ref{table-baseline-all}. The number of expected L and T dwarfs are shown in Figure~\ref{fig-numbers} using the \citet{2021AJ....161...42B} space densities. Note that the surveyed volume does not decrease monotonically with spectral type because early-T dwarfs are intrinsically brighter around one micron than late-L dwarfs.

\begin{deluxetable*}{lrrr}
\tablenum{1}
\tablecaption{V1.7 Baseline Survey Results \label{table-baseline-all}}
\tablewidth{0pt}
\tablehead{
\colhead{Type} & 
\colhead{Volume} & 
\colhead{Number} &
\colhead{Number}  \\
\colhead{} & 
\colhead{($10^5$ pc$^3$)} &
\colhead{(K21)} &
\colhead{(B21)} 
}
\startdata
L0 & 31.03 & 962 & 993 \\
L1 & 23.68 & 1492 & 1492 \\
L2 & 18.39 & 865 & 809 \\
L3 & 11.84 & 331 & 544 \\
L4 & 6.27 & 414 & 307 \\
L5 & 5.14 & 339 & 355 \\
L6 & 3.66 & 296 & 124 \\
L7 & 1.81 & 108 & 63 \\
L8 & 1.92 & 59 & 96 \\
L9 & 2.78 & 175 & 111 \\
T0 & 2.29 & 55 & 30 \\
T1 & 1.23 & 30 & 22 \\
T2 & 2.12 & 93 & 74 \\
T3 & 2.57 & 69 & 49 \\
T4 & 2.89 & 136 & 104 \\
T5 & 1.15 & 136 & 105 \\
T6 & 0.68 & 98 & 66 \\
T7 & 0.23 & 35 & 32 \\
T8 & 0.17 & 59 & 22 \\
T9 & 0.13 & 24 & \nodata \\
\enddata
\tablerefs{K21: \citep{2021ApJS..253....7K}.  B21: \citep{2021AJ....161...42B}}
\end{deluxetable*}

In comparison, \citet{2021A&A...649A...6G} estimate that in the GCNS parallax sample, L2 dwarfs are 50\% incomplete at $\sim40$ pc and L4 dwarfs are 50\% incomplete at $\sim 25$pc .  
We conclude that in the WFD region, LSST outperforms Gaia for L and T dwarfs both in terms of limiting distance and surveyed volume and the existing parallax sample for mid-Ls to mid-Ts. The $\sim100$ objects per spectral type for cool Ls and warm Ts is nearly an order of magnitude increase on existing work.
The volumes in Table~\ref{table-baseline-all} can be compared to an all-sky spherical 20pc sample, $0.335\times10^5$ pc$^3$.  
Because T7 and later dwarfs are very faint in even $z$ and $y$, LSST probes a smaller volume than targeted efforts aimed at the 20pc sample, so that the LSST parallax survey represents at best an incremental contribution to the sample of the coolest brown dwarfs.

\begin{figure}
\plotone{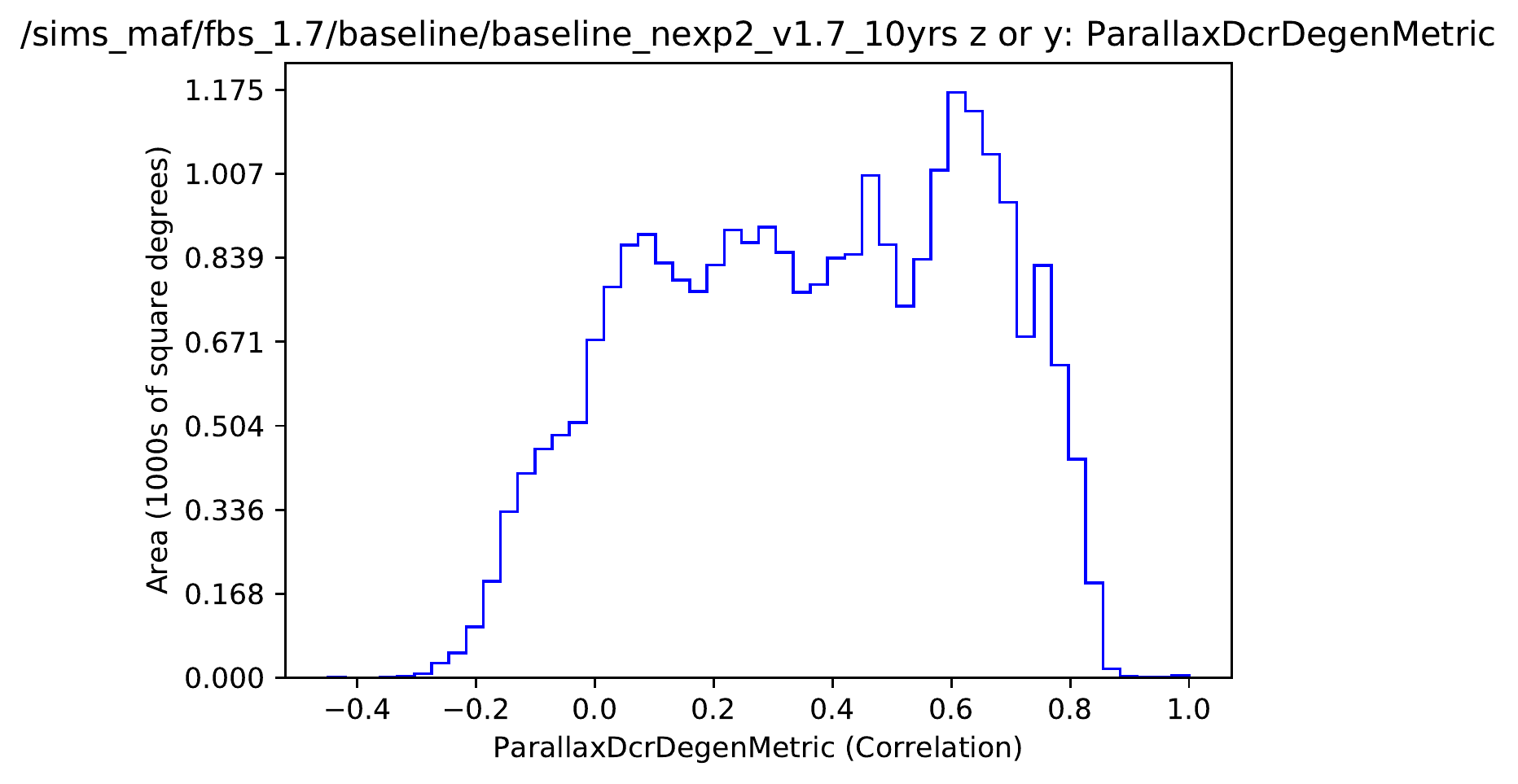}
\caption{\texttt{MAF} histogram of the correlation between the parallax signal and differential color refraction (DCR) due to the atmosphere. Only $z$ and $y$ observations are used. Cases where the correlation is greater than 0.8 are a concern for separating differential color refraction and the parallax signal for very red sources, but the simulations show this is a not a major problem for current survey strategies. \label{fig-dcr} }
\end{figure}

We also consider a metric checking on the correlation between the parallax factor and the DCR. If these are correlated (i.e., high parallax factors are observed when the DCR effect is large and in the same direction on the sky) then the solution becomes degenerate. (This effect is more important for the bluer $ugr$ filters.) \citet{2012AAS...21915607J} simulated LSST star and brown dwarf observations and found that parallaxes were unreliable when the correlation factor is greater than 0.8 (particularly if $ugr$ filters are used). However, the simulations show that the observations are uncorrelated over most of the sky (Figure~\ref{fig-dcr}). In the worst case scenario where all areas of the sky with correlation factor greater than 0.8 are excluded, the V1.7 baseline volume for L7 dwarfs is reduced by 1.3\% and the V2.0 baseline volume is reduced less than 0.2\%. We conclude that this effect is of little concern in selecting between the survey strategies currently under consideration.

\figsetstart
\figsetnum{2}
\figsettitle{Sky maps of distance limits}

\figsetgrpstart
\figsetgrpnum{2.1}
\figsetgrptitle{M6}
\figsetplot{fig2.M6.pdf}
\figsetgrpnote{The sky map for M6 dwarfs.}
\figsetgrpend

\figsetgrpstart
\figsetgrpnum{2.2}
\figsetgrptitle{M7}
\figsetplot{fig2.M7.pdf}
\figsetgrpnote{The sky map for M7 dwarfs.}
\figsetgrpend

\figsetgrpstart
\figsetgrpnum{2.3}
\figsetgrptitle{M8}
\figsetplot{fig2.M8.pdf}
\figsetgrpnote{The sky map for M8 dwarfs.}
\figsetgrpend

\figsetgrpstart
\figsetgrpnum{2.4}
\figsetgrptitle{M9}
\figsetplot{fig2.M9.pdf}
\figsetgrpnote{The sky map for M9 dwarfs.}
\figsetgrpend

\figsetgrpstart
\figsetgrpnum{2.5}
\figsetgrptitle{L0}
\figsetplot{fig2.L0.pdf}
\figsetgrpnote{The sky map for L0 dwarfs.}
\figsetgrpend

\figsetgrpstart
\figsetgrpnum{2.6}
\figsetgrptitle{L1}
\figsetplot{fig2.L1.pdf}
\figsetgrpnote{The sky map for L1 dwarfs.}
\figsetgrpend

\figsetgrpstart
\figsetgrpnum{2.7}
\figsetgrptitle{L2}
\figsetplot{fig2.L2.pdf}
\figsetgrpnote{The sky map for L2 dwarfs.}
\figsetgrpend

\figsetgrpstart
\figsetgrpnum{2.8}
\figsetgrptitle{L3}
\figsetplot{fig2.L3.pdf}
\figsetgrpnote{The sky map for L3 dwarfs.}
\figsetgrpend

\figsetgrpstart
\figsetgrpnum{2.9}
\figsetgrptitle{L4}
\figsetplot{fig2.L4.pdf}
\figsetgrpnote{The sky map for L4 dwarfs.}
\figsetgrpend

\figsetgrpstart
\figsetgrpnum{2.10}
\figsetgrptitle{L5}
\figsetplot{fig2.L5.pdf}
\figsetgrpnote{The sky map for L5 dwarfs.}
\figsetgrpend

\figsetgrpstart
\figsetgrpnum{2.11}
\figsetgrptitle{L6}
\figsetplot{fig2.L6.pdf}
\figsetgrpnote{The sky map for L6 dwarfs.}
\figsetgrpend

\figsetgrpstart
\figsetgrpnum{2.12}
\figsetgrptitle{L7}
\figsetplot{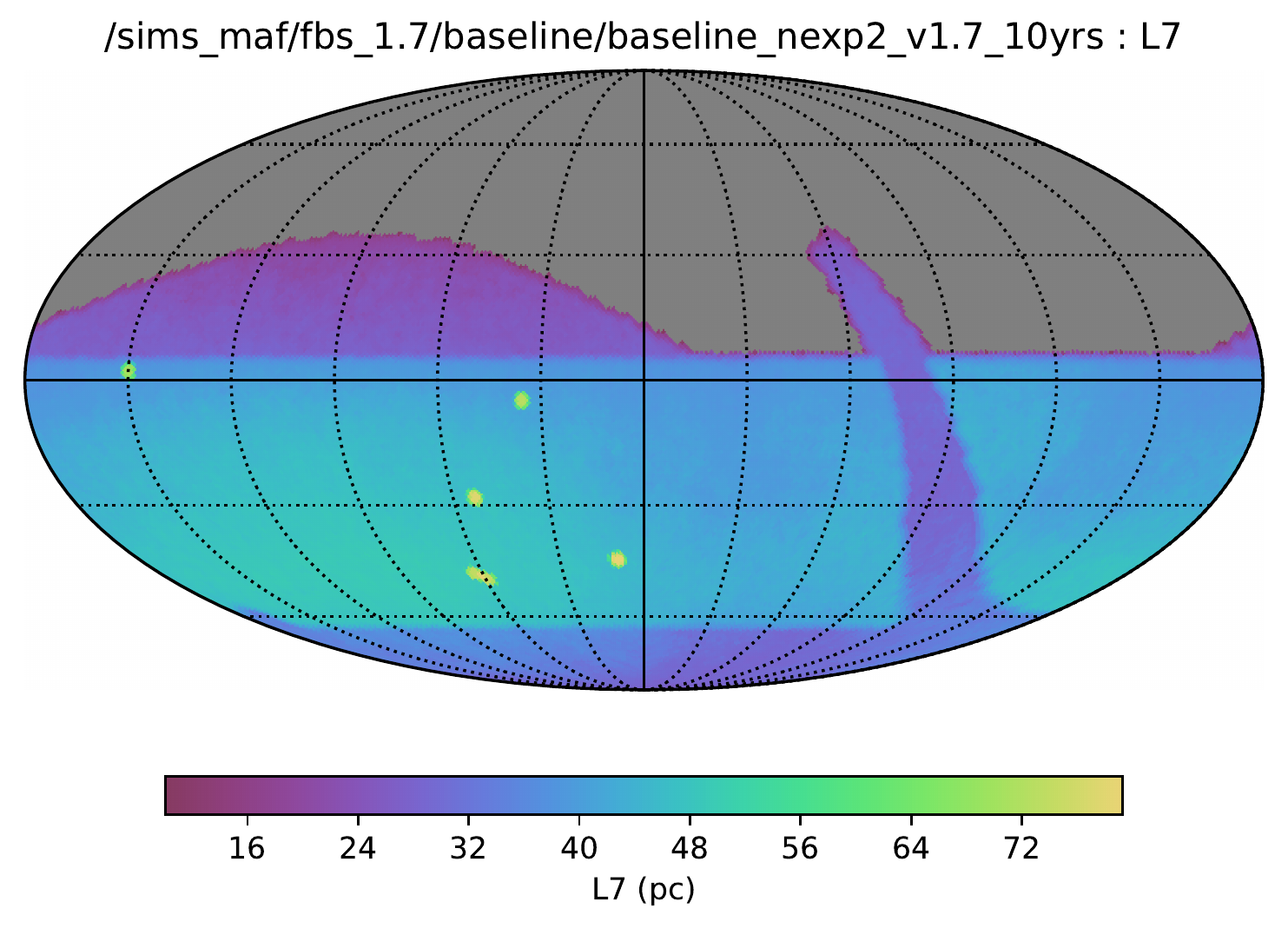}
\figsetgrpnote{The sky map for L7 dwarfs.}
\figsetgrpend

\figsetend

\figsetstart
\figsetnum{3}
\figsettitle{Histograms of distance limits}

\figsetgrpstart
\figsetgrpnum{3.1}
\figsetgrptitle{M6}
\figsetplot{fig3.M6.pdf}
\figsetgrpnote{The histogram for M6 dwarfs.}
\figsetgrpend

\figsetgrpstart
\figsetgrpnum{3.2}
\figsetgrptitle{M7}
\figsetplot{fig3.M7.pdf}
\figsetgrpnote{The histogram for M7 dwarfs.}
\figsetgrpend

\figsetgrpstart
\figsetgrpnum{3.3}
\figsetgrptitle{M8}
\figsetplot{fig3.M8.pdf}
\figsetgrpnote{The histogram for M8 dwarfs.}
\figsetgrpend

\figsetgrpstart
\figsetgrpnum{3.4}
\figsetgrptitle{M9}
\figsetplot{fig3.M9.pdf}
\figsetgrpnote{The histogram for M9 dwarfs.}
\figsetgrpend

\figsetgrpstart
\figsetgrpnum{3.5}
\figsetgrptitle{L0}
\figsetplot{fig3.L0.pdf}
\figsetgrpnote{The histogram for L0 dwarfs.}
\figsetgrpend

\figsetgrpstart
\figsetgrpnum{3.6}
\figsetgrptitle{L1}
\figsetplot{fig3.L1.pdf}
\figsetgrpnote{The histogram for L1 dwarfs.}
\figsetgrpend

\figsetgrpstart
\figsetgrpnum{3.7}
\figsetgrptitle{L2}
\figsetplot{fig3.L2.pdf}
\figsetgrpnote{The histogram for L2 dwarfs.}
\figsetgrpend

\figsetgrpstart
\figsetgrpnum{3.8}
\figsetgrptitle{L3}
\figsetplot{fig3.L3.pdf}
\figsetgrpnote{The histogram for L3 dwarfs.}
\figsetgrpend

\figsetgrpstart
\figsetgrpnum{3.9}
\figsetgrptitle{L4}
\figsetplot{fig3.L4.pdf}
\figsetgrpnote{The histogram for L4 dwarfs.}
\figsetgrpend

\figsetgrpstart
\figsetgrpnum{3.10}
\figsetgrptitle{L5}
\figsetplot{fig3.L5.pdf}
\figsetgrpnote{The histogram for L5 dwarfs.}
\figsetgrpend

\figsetgrpstart
\figsetgrpnum{3.11}
\figsetgrptitle{L6}
\figsetplot{fig3.L6.pdf}
\figsetgrpnote{The histogram for L6 dwarfs.}
\figsetgrpend

\figsetgrpstart
\figsetgrpnum{3.12}
\figsetgrptitle{L7}
\figsetplot{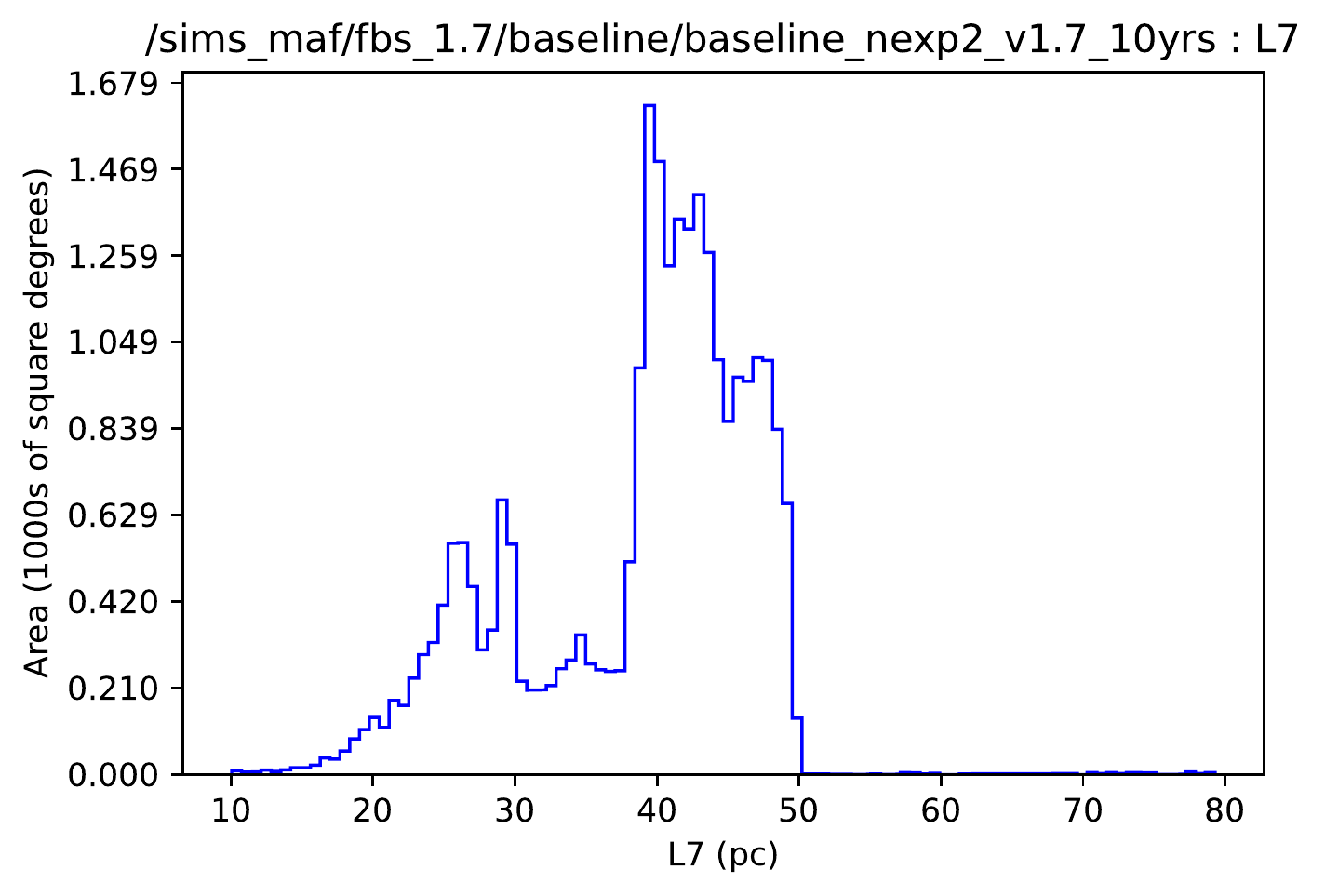}
\figsetgrpnote{The histogram for L7 dwarfs.}
\figsetgrpend

\figsetend

\section{Cadence Comparisons and Recommendations}

The initial set of \texttt{OpSim} simulations for this cadence optimization community process were dubbed V1.5, which included the baseline survey and various experiments in changing survey strategy. This V1.5 ``family" of simulated surveys was supplemented by additional experiments in the V1.6 and V1.7 families (see \citealt{jones_r_lynne_2020_4048838} for a guide to the simulations). Because the simulation assumptions, such as the seeing distribution, change significantly between the V1.5, V1.6 and V1.7 families, all simulations should be compared to their corresponding baseline survey in the same family. The simulations in the V1.5 and V1.6 families use only one 30s exposure per visit, but the 1.7 family returns to the original design of $2 \times 15$s exposures, which is less efficient and results in fewer total visits over the ten-year survey. We list the volumes probed by the baseline strategy in each family for both $1 \times 30$s and $2 \times 15$s in Table~\ref{table-compare-baselines}. In all families, the use of a single 30s exposure per visit improves the metrics  by $\sim 4-6$\% over the use of paired 15s exposures due to the greater number of visits. At a given number of exposures, volumes are larger in V1.5 and V1.6 primarily because of more favorable (but less realistic) assumptions. 

We compute the volume for L4 and L7 dwarfs for every \texttt{OpSim} simulation in the V1.5, V1.6, V1.7, and V1.7.1 families. To understand the effect of changes in the survey strategy, quantitative figures of merit compare each simulation to its corresponding baseline simulation.  We define the figure of merit as 

\begin{equation}
F_{\rm L4} = {{\rm Volume~for~L4~dwarfs}\over{\rm Baseline~Volume~for~L4~Dwarfs}}
\end{equation}

\begin{equation}
F_{\rm L7} = {{\rm Volume~for~L7~dwarfs}\over{\rm Baseline~Volume~for~L7~Dwarfs}}
\end{equation}

Values greater than one are desirable and those less than one are undesirable for this science case: A value of $F_{\rm L4} = 1.05$, for example, indicates that a simulation should have 5\% more L4 dwarfs with high quality parallaxes than the baseline. The L7 metric is the best representation of the relative performance around the critical L/T transition. L4 dwarfs are closer to the hydrogen-burning limit. They are also intrinsically brighter and are at a smaller parallax for a given limiting magnitude, and therefore are more sensitive to the parallax factor sampling. 

The complete set of metrics for all simulations is given in Table~\ref{table-metrics}. We note that all survey strategies aim to uniformly cover the WFD region in the first and last years of the survey in order to measure proper motions. We believe the brown dwarf parallax efforts also benefit because most strategies preferentially observe at low airmass, and hence near the meridian even at the start and end of the nights, and favor the use of $z$ or $y$ filters when sky backgrounds increase near twilight.

\begin{deluxetable*}{ccccl}
\tablenum{2}
\tablecaption{Baseline Survey Strategies\label{table-compare-baselines}}
\tablewidth{0pt}
\tablehead{
\colhead{Family} & 
\colhead{Exposure} & 
\colhead{L4 Volume} &
\colhead{L7 Volume} &
\colhead{Note} \\
\colhead{Version} & 
\colhead{(s)} & 
\colhead{($10^5$ pc$^3$)} &
\colhead{($10^5$ pc$^3$)} &
\colhead{}
}
\startdata
1.5 & $2\times15$ & 6.81 & 2.18 & \\
1.5 & $1\times30$ & 7.20 & 2.26 & V1.5 baseline \\
1.6 & $2\times15$ & 6.68 & 2.14 & \\
1.6 & $1\times30$ & 7.08 & 2.22 & V1.6 baseline \\
1.7 & $2\times15$ & 6.63 & 2.13 & V1.7 baseline \\
1.7 & $1\times30$ & 7.03 & 2.21 &  \\
2.0 & $2\times15$ & 6.19 & 2.05 & V2.0 baseline \\
\enddata
\end{deluxetable*}

The metrics in Table~\ref{table-metrics} generally vary by only a few percent, indicating that most proposed changes to the survey strategy have only minor effects on this science case, and many factors can trade off each other. The tradeoff is clearly revealed in the V1.5 \texttt{wfd\_depth} experiments without DDFs which split time between WFD and the mini-surveys.  In this particular set of simulations, the share of observing time devoted to WFD ranges from 99\% down to 65\%; the remaining 1\% to 35\% is devoted to the mini-surveys including the Galactic plane, and 0\% is devoted to DDF fields.  
Here the metrics change by only 1-2\% as the WFD share drops from 99\% to 65\%; the loss in WFD depth is compensated by the gain in area and depth elsewhere, with the L4 metric preferring deeper WFD and the L7 metric preferring shallower WFD but deeper mini-surveys. Amongst the full range of simulations, many of the most extreme results, like the alternative footprint \texttt{newA} ($F_{\rm L7} = 1.10$) or the stuck rolling cadence ($F_{\rm L7} = 0.84$) were already known to violate the SRD requirements. 
We can see, however, that that the simulations are consistent with expectations by examining limiting cases. One way to improve the brown dwarf parallax sample would be to simply observe more in the $z$ and $y$ bands instead of the bluer bands, which are not used for this particular science case. The V1.5 \texttt{filt\_dist} experiments \texttt{filterdist\_indx5} ($F_{\rm L4} =1.19$,$F_{\rm L7}=1.10$)  and \texttt{filterdist\_indx8} ($F_{\rm L4} =1.08$,$F_{\rm L7}=1.03$) show that greatly increasing the number of $z$ and $y$ visits improves the metrics, while the experiments reducing $z$ and $y$ visits in turn degrade the metric. 

Many of the proposed changes have little impact on our science case, changing the volume by $\sim \pm1$\%.  Examples include the \texttt{even\_filters} and \texttt{wfd\_cadence\_drive} experiments which try to reduce the baseline simulation's tendency to observe
with the reddest filters in bright time and the bluer filters in dark time, experiments related to how deep drilling fields are dithered (\texttt{ddf\_dither, euclid\_dither}, experiments on making twilight observations paired in the same or different filter (\texttt{twi\_pairs}), and experiments including occasional high airmass observations in the bluer filters \texttt{dcr}.  Evidently these choices have little effect on the $i,z, y$ sampling of the parallax signal over a decade. Varying the observing time (\texttt{var\_expt}) between 20s and 100s to make the image depth more uniform, instead of always using 30s, also has little effect even though it reduces the total number of visits.

As one would expect, strategies that involve reducing the number of $izy$ WFD observations to enable other types of observations potentially have negative impact on this science case, but most of the ones simulated have only modest (few percent) effects. Examples include devoting time to conduct occasional short exposure (\texttt{short\_exp} surveys and increasing the total time observing with the $u$ filter (\texttt{u\_long}). Similarly, surveys that favor using bluer filters when the best seeing conditions occur (\texttt{goodseeing}) has a small negative effect because astrometry does benefit from good seeing in $y$.

We identify tension with other uses of twilight time which would affect the parallax sampling: An extreme example is the nightly Near-Earth Object (NEO) search (\texttt{twi\_neo pattern1\_v1.7\_10yrs}, which has a large negative impact ($F_{\rm L4}=0.90$,$F_{\rm L7}=0.94$). The simulated NEO searches use short (1s) $riz$ exposures that are not suitable for precision astrometry of faint brown dwarfs. \citet{jones_r_lynne_2020_4048838} note that given the weather assumptions, some twilight time must be used for WFD to meet SRD coverage requirements,  which rules out executing the NEO search every night. In V1.7, NEO Patterns 3, 4 and 7 have the smallest impact on our science case and would be most acceptable. 

One possible strategy is some form of ``rolling cadence" scheduling.This would observed selected regions of the sky more intensely in certain years, but less intensely in most years, for the same total number of observations. In the V1.7 family, the rolling cadence (\texttt{rolling}, \texttt{rolling\_nm}) have no significant effect. 

The remaining major science strategy issues are the WFD footprint on the sky and amount of time of devoted to WFD and mini-surveys, which can have the largest impact on the metrics for this science case. Relevant simulations in Table~\ref{table-metrics} include the \texttt{bulge} simulations that add mini-surveys in the Galactic bulge and plane, \texttt{potential\_schedulers},and \texttt{footprints}. We discuss the many possibilities in the next section where we include our responses and recommendations in response to specific SCOC questions. 

Before addressing the specific questions asked by the SCOC, we make two recommendations. First, we recommend the single exposure per visit because gain for a single 30s snap over 2x15s is 4-6\% (Table~\ref{table-compare-baselines}) in all families, enabling better parallax factor sampling in $z,y$ or more mini-survey coverage. The choice of one or two exposures will be studied in commissioning \citep{2019ApJ...873..111I}. We also recommend studying astrometric performance and software in commissioning. The 10 mas atmospheric and instrumental astrometric uncertainty goal may be overly conservative now that Gaia provides a dense ``truth" sample in every image. \citet{2017PASP..129g4503B} and \citet{2021AJ....162..106F} show how novel algorithms applied to the Gaia reference stars in Dark Energy Survey data enable improvement of camera distortions and atmospheric turbulence residuals from 10 mas to 5-7 mas. As an ideal case, we reran the \texttt{MAF} simulations with 5 mas uncertainty instead of 10 mas -- this improves the metrics for V1.7 baseline simulations by 7-15\% without any changes to survey strategy.

\subsection{SCOC Questions}

The specific questions asked by the SCOC for the Cadence Notes are given in Appendix B.3 of \citet{2022ApJS..258....1B}. Here we describe our answers to these questions. 

\textit{Questions 1 and 2: Are there any science drivers that would strongly argue for, or against, increasing the WFD footprint from 18,000
square degrees to 20,000 square degrees? Note that the resulting number of visits would drop by about 10\%. Assuming current system performance estimates will hold up, we plan to utililze the additional observing time for visits for the mini-surveys and the DDFs. What is the best scientific use of this time?}

These questions address major issues in the LSST survey design.  The V1.5, V1.6, and V1.7 baselines all meet the minimum WFD survey requirement of 18,000 degrees but often exceed the minimum requirement of 825 visits. One can imagine covering more area with fewer visits, or even shifting excess visits to other parts of the sky. The footprint on the sky could also be shifted. The baseline simulations include 5\% of the time spent on DDFs and limited observations of Galactic bulge and central Galactic plane, Magellanic Clouds, South Celestial Cap, and North Ecliptic Spur.

We find that surveys that skimp on WFD visits result in poorer parallaxes. However, the tradeoff between sky coverage and WFD depth is complex. As discussed above, trading off area and WFD depth can largely cancel out in the brown dwarf parallax metrics. 
However, we see that simulations without any low galactic latitude coverage and/or minimum coverage in WFD have very large negative impacts (examples: V1.5 \texttt{big\_sky}, \texttt{big\_sky\_no\_uiy}, \texttt{big\_sky\_dust}, V1.6 \texttt{ddf\_heavy}) and we recommend against them. The Big Sky concept moves WFD north and south and excludes even the outer Galactic plane to avoid dust extinctions, but in these particular experiments have no low Galactic latitude observations at all. WFD larger footprints and mini-surveys that increase the coverage at low galactic latitude or the northern hemisphere can have significant improvements, such as (\texttt{bulges} family, \texttt{newA}, \texttt{newB}); \texttt{combo\_dust} is a good V1.6 potential scheduler, which increases the total area over an adjusted footprint but at the cost of fewer visits. The \texttt{new\_rolling} (V1.17.1) cases with plane coverage have small but positive impacts. We emphasize the illustrative contrast between the very poor parallax performance of the proposed Big Sky strategy \texttt{big\_skyv1.5} ($F_{\rm L4}=0.88$, $F_{\rm L7}=0.90$) without plane coverage, with Big Sky at a standard WFD cadence including low galactic latitudes \texttt{big\_sky\_wfd1.5} (FL4=1.02, FL7 =1.06).  (Unfortunately, while this science case would benefit from newA or Big Sky WFD, they ultimately require too much observing time and do not meet the SRD requirements.) Brown dwarf astrometry disfavors adding more pencil beam surveys in the form of DDFs (\texttt{wfd\_depth} family, V1.6 \texttt{ddf\_heavy}) at the expense of WFD or mini-surveys with extended area. Most small adjustments, such as the V1.7 \texttt{footprint\_tune} family, have no importance to this science case, with changes of $\sim1$\% over baseline. 

\textit{Question 3: Are there any science drivers that would strongly argue for, or against, the proposal to change the $u$-band exposure from $2 \times 15$ s to $1 \times 50$ s?}

This science case makes no use of $u$-band, but it is important to other SMWLV science cases. Changing $u$-band to 50s has a modest negative impact (\texttt{u\_long\_50}) of 2\% due to the corresponding reduction in red filter observations, but may have positive effects on other SMWLV science cases. 

\textit{Question 4: Are there any science drivers that would strongly argue for, or against, further changes in observing time allocation per band (e.g., skewed much more toward
the blue or the red side of the spectrum)?}

This science case gains when skewed much more to the red (V1.5 \texttt{filter\_dist}) but loses when skewed much more to the blue (V1.5 \texttt{bluer\_footprint}, \texttt{filter\_dist} index 1). We see red filter reductions as highly undesirable and strongly recommend against them. 

\textit{Question 5:
Are there any science drivers that would strongly argue
for, or against, obtaining two visits in a pair in the same (or
different) filter? Or the benefits or drawbacks of dedicating
a portion of each night to obtaining a third (triplet) visit? }

The pairs of visits do not affect our science case. Reserving the end of the night (when parallax factors are high) for third observations is a potential concern. However, we find no impact when 30 minutes (V1.5 \texttt{third\_obs}) is reserved. We do see small negative impacts as the reserved time increases, reaching 1\% in the 120 minute case. Strategies (V1.5 \texttt{good\_seeing}) that select against quality $y$-band observations are negative at the few percent level.

\textit{Questions 6 and 7: Are there any science drivers that would strongly argue for, or against, the rolling-cadence scenario? Are there any science drivers pushing for or against particular
dithering patterns (either rotational dithers or translational
dithers?}

We find little effect for rolling cadence (in V1.6 or V1.7). The now obsolete rolling cadence algorithm as implemented in V1.5 does have negative impact, so this should be monitored. Our model does not capture any effects from the dithering simulations, though we remark dithering is desirable for astrometric measurements. The PSF is expected to be well sampled in most cases with the camera scale of 0.2 arcseconds per pixel. 

\startlongtable
\begin{deluxetable*}{llll}
\tablenum{3}
\tablecaption{Metrics\label{table-metrics}}
\tablewidth{0pt}
\tablehead{
\colhead{$F_{\rm L4}$} &
\colhead{$F_{\rm L7}$} &
\colhead{Family} &
\colhead{Simulation Name} 
}
\startdata
0.989 & 1.032 & 1.5 & alt\_roll\_dust/alt\_dust\_v1.5\_10yrs.db \\
0.967 & 1.025 & 1.5 & alt\_roll\_dust/alt\_roll\_mod2\_dust\_sdf\_0.20\_v1.5\_10yrs.db \\
0.974 & 1.031 & 1.5 & alt\_roll\_dust/roll\_mod2\_dust\_sdf\_0.20\_v1.5\_10yrs.db \\
0.946 & 0.965 & 1.5 & baseline/baseline\_2snaps\_v1.5\_10yrs.db \\
1.000 & 1.000 & 1.5 & baseline/baseline\_v1.5\_10yrs.db \\
1.001 & 1.047 & 1.5 & bulge/bulges\_bs\_v1.5\_10yrs.db \\
1.007 & 1.054 & 1.5 & bulge/bulges\_bulge\_wfd\_v1.5\_10yrs.db \\
1.000 & 1.046 & 1.5 & bulge/bulges\_cadence\_bs\_v1.5\_10yrs.db \\
1.005 & 1.052 & 1.5 & bulge/bulges\_cadence\_bulge\_wfd\_v1.5\_10yrs.db \\
1.002 & 1.051 & 1.5 & bulge/bulges\_cadence\_i\_heavy\_v1.5\_10yrs.db \\
1.002 & 1.051 & 1.5 & bulge/bulges\_i\_heavy\_v1.5\_10yrs.db \\
0.965 & 0.977 & 1.5 & daily\_ddf/daily\_ddf\_v1.5\_10yrs.db \\
0.988 & 0.990 & 1.5 & dcr/dcr\_nham1\_ug\_v1.5\_10yrs.db \\
0.993 & 0.996 & 1.5 & dcr/dcr\_nham1\_ugr\_v1.5\_10yrs.db \\
0.989 & 0.992 & 1.5 & dcr/dcr\_nham1\_ugri\_v1.5\_10yrs.db \\
0.990 & 0.992 & 1.5 & dcr/dcr\_nham2\_ug\_v1.5\_10yrs.db \\
0.989 & 0.992 & 1.5 & dcr/dcr\_nham2\_ugr\_v1.5\_10yrs.db \\
0.981 & 0.991 & 1.5 & dcr/dcr\_nham2\_ugri\_v1.5\_10yrs.db \\
0.911 & 0.919 & 1.5 & filter\_dist/filterdist\_indx1\_v1.5\_10yrs.db \\
1.022 & 0.988 & 1.5 & filter\_dist/filterdist\_indx2\_v1.5\_10yrs.db \\
0.936 & 0.935 & 1.5 & filter\_dist/filterdist\_indx3\_v1.5\_10yrs.db \\
0.954 & 0.947 & 1.5 & filter\_dist/filterdist\_indx4\_v1.5\_10yrs.db \\
1.194 & 1.100 & 1.5 & filter\_dist/filterdist\_indx5\_v1.5\_10yrs.db \\
0.978 & 0.956 & 1.5 & filter\_dist/filterdist\_indx6\_v1.5\_10yrs.db \\
0.973 & 0.958 & 1.5 & filter\_dist/filterdist\_indx7\_v1.5\_10yrs.db \\
1.084 & 1.029 & 1.5 & filter\_dist/filterdist\_indx8\_v1.5\_10yrs.db \\
1.000 & 1.003 & 1.5 & footprints/footprint\_add\_mag\_cloudsv1.5\_10yrs.db \\
0.938 & 0.949 & 1.5 & footprints/footprint\_big\_sky\_dustv1.5\_10yrs.db \\
0.901 & 0.913 & 1.5 & footprints/footprint\_big\_sky\_nouiyv1.5\_10yrs.db \\
0.888 & 0.907 & 1.5 & footprints/footprint\_big\_skyv1.5\_10yrs.db \\
1.024 & 1.060 & 1.5 & footprints/footprint\_big\_wfdv1.5\_10yrs.db \\
0.864 & 0.901 & 1.5 & footprints/footprint\_bluer\_footprintv1.5\_10yrs.db \\
1.018 & 1.029 & 1.5 & footprints/footprint\_gp\_smoothv1.5\_10yrs.db \\
1.051 & 1.095 & 1.5 & footprints/footprint\_newAv1.5\_10yrs.db \\
1.008 & 1.056 & 1.5 & footprints/footprint\_newBv1.5\_10yrs.db \\
0.995 & 0.991 & 1.5 & footprints/footprint\_no\_gp\_northv1.5\_10yrs.db \\
0.999 & 0.999 & 1.5 & footprints/footprint\_standard\_goalsv1.5\_10yrs.db \\
0.838 & 0.837 & 1.5 & footprints/footprint\_stuck\_rollingv1.5\_10yrs.db \\
0.971 & 0.978 & 1.5 & goodseeing/goodseeing\_gi\_v1.5\_10yrs.db \\
0.970 & 0.976 & 1.5 & goodseeing/goodseeing\_gri\_v1.5\_10yrs.db \\
0.984 & 0.989 & 1.5 & goodseeing/goodseeing\_griz\_v1.5\_10yrs.db \\
0.981 & 0.985 & 1.5 & goodseeing/goodseeing\_gz\_v1.5\_10yrs.db \\
0.986 & 0.987 & 1.5 & goodseeing/goodseeing\_i\_v1.5\_10yrs.db \\
0.987 & 0.993 & 1.5 & greedy\_footprint/greedy\_footprint\_v1.5\_10yrs.db \\
0.965 & 0.980 & 1.5 & rolling/rolling\_mod2\_sdf\_0.10\_v1.5\_10yrs.db \\
0.960 & 0.975 & 1.5 & rolling/rolling\_mod2\_sdf\_0.20\_v1.5\_10yrs.db \\
0.938 & 0.964 & 1.5 & rolling/rolling\_mod3\_sdf\_0.10\_v1.5\_10yrs.db \\
0.940 & 0.965 & 1.5 & rolling/rolling\_mod3\_sdf\_0.20\_v1.5\_10yrs.db \\
0.914 & 0.950 & 1.5 & rolling/rolling\_mod6\_sdf\_0.10\_v1.5\_10yrs.db \\
0.920 & 0.954 & 1.5 & rolling/rolling\_mod6\_sdf\_0.20\_v1.5\_10yrs.db \\
0.977 & 0.982 & 1.5 & same\_filt/baseline\_samefilt\_v1.5\_10yrs.db \\
0.992 & 0.988 & 1.5 & short\_exp/short\_exp\_2ns\_1expt\_v1.5\_10yrs.db \\
0.985 & 0.983 & 1.5 & short\_exp/short\_exp\_2ns\_5expt\_v1.5\_10yrs.db \\
0.984 & 0.980 & 1.5 & short\_exp/short\_exp\_5ns\_1expt\_v1.5\_10yrs.db \\
0.964 & 0.968 & 1.5 & short\_exp/short\_exp\_5ns\_5expt\_v1.5\_10yrs.db \\
1.003 & 1.001 & 1.5 & spiders/spiders\_v1.5\_10yrs.db \\
0.988 & 0.990 & 1.5 & third\_obs/third\_obs\_pt120v1.5\_10yrs.db \\
1.000 & 1.000 & 1.5 & third\_obs/third\_obs\_pt15v1.5\_10yrs.db \\
1.001 & 1.000 & 1.5 & third\_obs/third\_obs\_pt30v1.5\_10yrs.db \\
1.000 & 1.000 & 1.5 & third\_obs/third\_obs\_pt45v1.5\_10yrs.db \\
0.995 & 0.995 & 1.5 & third\_obs/third\_obs\_pt60v1.5\_10yrs.db \\
0.994 & 0.995 & 1.5 & third\_obs/third\_obs\_pt90v1.5\_10yrs.db \\
0.979 & 1.012 & 1.5 & twilight\_neo/twilight\_neo\_mod1\_v1.5\_10yrs.db \\
0.995 & 1.014 & 1.5 & twilight\_neo/twilight\_neo\_mod2\_v1.5\_10yrs.db \\
0.993 & 1.007 & 1.5 & twilight\_neo/twilight\_neo\_mod3\_v1.5\_10yrs.db \\
0.992 & 1.003 & 1.5 & twilight\_neo/twilight\_neo\_mod4\_v1.5\_10yrs.db \\
0.980 & 0.987 & 1.5 & u60/u60\_v1.5\_10yrs.db \\
0.992 & 0.999 & 1.5 & var\_expt/var\_expt\_v1.5\_10yrs.db \\
0.998 & 1.025 & 1.5 & wfd\_depth/wfd\_depth\_scale0.65\_noddf\_v1.5\_10yrs.db \\
0.974 & 1.006 & 1.5 & wfd\_depth/wfd\_depth\_scale0.65\_v1.5\_10yrs.db \\
1.002 & 1.023 & 1.5 & wfd\_depth/wfd\_depth\_scale0.70\_noddf\_v1.5\_10yrs.db \\
0.978 & 1.005 & 1.5 & wfd\_depth/wfd\_depth\_scale0.70\_v1.5\_10yrs.db \\
1.009 & 1.021 & 1.5 & wfd\_depth/wfd\_depth\_scale0.75\_noddf\_v1.5\_10yrs.db \\
0.986 & 1.004 & 1.5 & wfd\_depth/wfd\_depth\_scale0.75\_v1.5\_10yrs.db \\
1.013 & 1.011 & 1.5 & wfd\_depth/wfd\_depth\_scale0.80\_noddf\_v1.5\_10yrs.db \\
0.992 & 0.998 & 1.5 & wfd\_depth/wfd\_depth\_scale0.80\_v1.5\_10yrs.db \\
1.020 & 1.009 & 1.5 & wfd\_depth/wfd\_depth\_scale0.85\_noddf\_v1.5\_10yrs.db \\
0.995 & 0.996 & 1.5 & wfd\_depth/wfd\_depth\_scale0.85\_v1.5\_10yrs.db \\
1.019 & 1.007 & 1.5 & wfd\_depth/wfd\_depth\_scale0.90\_noddf\_v1.5\_10yrs.db \\
1.000 & 1.000 & 1.5 & wfd\_depth/wfd\_depth\_scale0.90\_v1.5\_10yrs.db \\
1.017 & 1.003 & 1.5 & wfd\_depth/wfd\_depth\_scale0.95\_noddf\_v1.5\_10yrs.db \\
0.996 & 0.988 & 1.5 & wfd\_depth/wfd\_depth\_scale0.95\_v1.5\_10yrs.db \\
1.023 & 1.002 & 1.5 & wfd\_depth/wfd\_depth\_scale0.99\_noddf\_v1.5\_10yrs.db \\
1.003 & 0.990 & 1.5 & wfd\_depth/wfd\_depth\_scale0.99\_v1.5\_10yrs.db\\
1.003 & 1.005 & 1.6 & even\_filters/even\_filters\_alt\_g\_v1.6\_10yrs.db \\
0.999 & 1.000 & 1.6 & even\_filters/even\_filters\_altv1.6\_10yrs.db \\
1.009 & 1.008 & 1.6 & even\_filters/even\_filters\_g\_v1.6\_10yrs.db \\
1.004 & 1.006 & 1.6 & even\_filters/even\_filtersv1.6\_10yrs.db \\
0.941 & 0.920 & 1.6 & potential\_schedulers/barebones\_nexp2\_v1.6\_10yrs.db \\
0.997 & 0.958 & 1.6 & potential\_schedulers/barebones\_v1.6\_10yrs.db \\
1.000 & 1.000 & 1.6 & potential\_schedulers/baseline\_nexp1\_v1.6\_10yrs.db \\
0.937 & 0.955 & 1.6 & potential\_schedulers/baseline\_nexp2\_scaleddown\_v1.6\_10yrs.db \\
0.944 & 0.963 & 1.6 & potential\_schedulers/baseline\_nexp2\_v1.6\_10yrs.db \\
0.948 & 0.995 & 1.6 & potential\_schedulers/combo\_dust\_nexp2\_v1.6\_10yrs.db \\
1.002 & 1.038 & 1.6 & potential\_schedulers/combo\_dust\_v1.6\_10yrs.db \\
0.883 & 0.923 & 1.6 & potential\_schedulers/ddf\_heavy\_nexp2\_v1.6\_10yrs.db \\
0.939 & 0.963 & 1.6 & potential\_schedulers/ddf\_heavy\_v1.6\_10yrs.db \\
0.913 & 0.935 & 1.6 & potential\_schedulers/dm\_heavy\_nexp2\_v1.6\_10yrs.db \\
0.963 & 0.970 & 1.6 & potential\_schedulers/dm\_heavy\_v1.6\_10yrs.db \\
0.949 & 0.974 & 1.6 & potential\_schedulers/mw\_heavy\_nexp2\_v1.6\_10yrs.db \\
1.002 & 1.008 & 1.6 & potential\_schedulers/mw\_heavy\_v1.6\_10yrs.db \\
0.932 & 0.973 & 1.6 & potential\_schedulers/rolling\_exgal\_mod2\_dust\_sdf\_0.80\_nexp2\_v1.6\_10yrs.db \\
0.982 & 1.014 & 1.6 & potential\_schedulers/rolling\_exgal\_mod2\_dust\_sdf\_0.80\_v1.6\_10yrs.db \\
0.930 & 0.965 & 1.6 & potential\_schedulers/ss\_heavy\_nexp2\_v1.6\_10yrs.db \\
0.991 & 1.007 & 1.6 & potential\_schedulers/ss\_heavy\_v1.6\_10yrs.db \\
1.001 & 1.002 & 1.6 & rolling\_fpo/rolling\_fpo\_2nslice0.8\_v1.6\_10yrs.db \\
1.002 & 1.003 & 1.6 & rolling\_fpo/rolling\_fpo\_2nslice0.9\_v1.6\_10yrs.db \\
1.000 & 1.002 & 1.6 & rolling\_fpo/rolling\_fpo\_2nslice1.0\_v1.6\_10yrs.db \\
0.990 & 0.990 & 1.6 & rolling\_fpo/rolling\_fpo\_3nslice0.8\_v1.6\_10yrs.db \\
0.990 & 0.990 & 1.6 & rolling\_fpo/rolling\_fpo\_3nslice0.9\_v1.6\_10yrs.db \\
0.989 & 0.987 & 1.6 & rolling\_fpo/rolling\_fpo\_3nslice1.0\_v1.6\_10yrs.db \\
0.998 & 0.994 & 1.6 & rolling\_fpo/rolling\_fpo\_6nslice0.8\_v1.6\_10yrs.db \\
0.995 & 0.993 & 1.6 & rolling\_fpo/rolling\_fpo\_6nslice0.9\_v1.6\_10yrs.db \\
0.998 & 0.994 & 1.6 & rolling\_fpo/rolling\_fpo\_6nslice1.0\_v1.6\_10yrs.db \\
1.060 & 1.040 & 1.7 & baseline/baseline\_nexp1\_v1.7\_10yrs.db \\
1.000 & 1.000 & 1.7 & baseline/baseline\_nexp2\_v1.7\_10yrs.db \\
1.000 & 1.001 & 1.7 & ddf\_dither/ddf\_dither0.00\_v1.7\_10yrs.db \\
1.000 & 1.000 & 1.7 & ddf\_dither/ddf\_dither0.05\_v1.7\_10yrs.db \\
1.000 & 1.000 & 1.7 & ddf\_dither/ddf\_dither0.10\_v1.7\_10yrs.db \\
1.002 & 1.001 & 1.7 & ddf\_dither/ddf\_dither0.30\_v1.7\_10yrs.db \\
1.000 & 1.000 & 1.7 & ddf\_dither/ddf\_dither0.70\_v1.7\_10yrs.db \\
1.002 & 1.001 & 1.7 & ddf\_dither/ddf\_dither1.00\_v1.7\_10yrs.db \\
1.003 & 1.001 & 1.7 & ddf\_dither/ddf\_dither1.50\_v1.7\_10yrs.db \\
1.002 & 1.000 & 1.7 & ddf\_dither/ddf\_dither2.00\_v1.7\_10yrs.db \\
1.000 & 1.000 & 1.7 & euclid\_dither/euclid\_dither1\_v1.7\_10yrs.db \\
1.002 & 1.001 & 1.7 & euclid\_dither/euclid\_dither2\_v1.7\_10yrs.db \\
1.000 & 1.000 & 1.7 & euclid\_dither/euclid\_dither3\_v1.7\_10yrs.db \\
0.999 & 1.000 & 1.7 & euclid\_dither/euclid\_dither4\_v1.7\_10yrs.db \\
1.000 & 1.000 & 1.7 & euclid\_dither/euclid\_dither5\_v1.7\_10yrs.db \\
0.992 & 1.013 & 1.7 & footprint\_tune/footprint\_0\_v1.710yrs.db \\
0.995 & 1.014 & 1.7 & footprint\_tune/footprint\_1\_v1.710yrs.db \\
0.991 & 1.010 & 1.7 & footprint\_tune/footprint\_2\_v1.710yrs.db \\
0.995 & 1.012 & 1.7 & footprint\_tune/footprint\_3\_v1.710yrs.db \\
0.998 & 1.003 & 1.7 & footprint\_tune/footprint\_4\_v1.710yrs.db \\
0.995 & 1.012 & 1.7 & footprint\_tune/footprint\_5\_v1.710yrs.db \\
0.997 & 1.010 & 1.7 & footprint\_tune/footprint\_6\_v1.710yrs.db \\
0.991 & 1.008 & 1.7 & footprint\_tune/footprint\_7\_v1.710yrs.db \\
0.982 & 1.003 & 1.7 & footprint\_tune/footprint\_8\_v1.710yrs.db \\
1.005 & 1.006 & 1.7 & pair\_times/pair\_times\_11\_v1.7\_10yrs.db \\
1.000 & 1.000 & 1.7 & pair\_times/pair\_times\_22\_v1.7\_10yrs.db \\
0.994 & 0.994 & 1.7 & pair\_times/pair\_times\_33\_v1.7\_10yrs.db \\
0.984 & 0.989 & 1.7 & pair\_times/pair\_times\_44\_v1.7\_10yrs.db \\
0.970 & 0.978 & 1.7 & pair\_times/pair\_times\_55\_v1.7\_10yrs.db \\
1.002 & 1.001 & 1.7 & rolling/rolling\_scale0.2\_nslice2\_v1.7\_10yrs.db \\
1.000 & 1.001 & 1.7 & rolling/rolling\_scale0.2\_nslice3\_v1.7\_10yrs.db \\
1.000 & 1.000 & 1.7 & rolling/rolling\_scale0.4\_nslice2\_v1.7\_10yrs.db \\
1.003 & 1.002 & 1.7 & rolling/rolling\_scale0.4\_nslice3\_v1.7\_10yrs.db \\
1.001 & 1.000 & 1.7 & rolling/rolling\_scale0.6\_nslice2\_v1.7\_10yrs.db \\
1.003 & 1.003 & 1.7 & rolling/rolling\_scale0.6\_nslice3\_v1.7\_10yrs.db \\
1.002 & 1.002 & 1.7 & rolling/rolling\_scale0.8\_nslice2\_v1.7\_10yrs.db \\
1.006 & 1.007 & 1.7 & rolling/rolling\_scale0.8\_nslice3\_v1.7\_10yrs.db \\
1.000 & 1.001 & 1.7 & rolling/rolling\_scale0.9\_nslice2\_v1.7\_10yrs.db \\
1.005 & 1.006 & 1.7 & rolling/rolling\_scale0.9\_nslice3\_v1.7\_10yrs.db \\
1.000 & 1.001 & 1.7 & rolling/rolling\_scale1.0\_nslice2\_v1.7\_10yrs.db \\
1.005 & 1.006 & 1.7 & rolling/rolling\_scale1.0\_nslice3\_v1.7\_10yrs.db \\
0.998 & 1.000 & 1.7 & rolling\_nm/rolling\_nm\_scale0.2\_nslice2\_v1.7\_10yrs.db \\
0.998 & 1.000 & 1.7 & rolling\_nm/rolling\_nm\_scale0.2\_nslice3\_v1.7\_10yrs.db \\
0.998 & 1.000 & 1.7 & rolling\_nm/rolling\_nm\_scale0.4\_nslice2\_v1.7\_10yrs.db \\
1.004 & 1.003 & 1.7 & rolling\_nm/rolling\_nm\_scale0.4\_nslice3\_v1.7\_10yrs.db \\
1.003 & 1.004 & 1.7 & rolling\_nm/rolling\_nm\_scale0.6\_nslice2\_v1.7\_10yrs.db \\
1.006 & 1.006 & 1.7 & rolling\_nm/rolling\_nm\_scale0.6\_nslice3\_v1.7\_10yrs.db \\
1.002 & 1.004 & 1.7 & rolling\_nm/rolling\_nm\_scale0.8\_nslice2\_v1.7\_10yrs.db \\
1.009 & 1.010 & 1.7 & rolling\_nm/rolling\_nm\_scale0.8\_nslice3\_v1.7\_10yrs.db \\
1.003 & 1.005 & 1.7 & rolling\_nm/rolling\_nm\_scale0.9\_nslice2\_v1.7\_10yrs.db \\
1.006 & 1.007 & 1.7 & rolling\_nm/rolling\_nm\_scale0.9\_nslice3\_v1.7\_10yrs.db \\
1.003 & 1.006 & 1.7 & rolling\_nm/rolling\_nm\_scale1.0\_nslice2\_v1.7\_10yrs.db \\
1.011 & 1.013 & 1.7 & rolling\_nm/rolling\_nm\_scale1.0\_nslice3\_v1.7\_10yrs.db \\
0.904 & 0.941 & 1.7 & twi\_neo/twi\_neo\_pattern1\_v1.7\_10yrs.db \\
0.953 & 0.972 & 1.7 & twi\_neo/twi\_neo\_pattern2\_v1.7\_10yrs.db \\
0.971 & 0.984 & 1.7 & twi\_neo/twi\_neo\_pattern3\_v1.7\_10yrs.db \\
0.980 & 0.987 & 1.7 & twi\_neo/twi\_neo\_pattern4\_v1.7\_10yrs.db \\
0.956 & 0.974 & 1.7 & twi\_neo/twi\_neo\_pattern5\_v1.7\_10yrs.db \\
0.960 & 0.976 & 1.7 & twi\_neo/twi\_neo\_pattern6\_v1.7\_10yrs.db \\
0.970 & 0.981 & 1.7 & twi\_neo/twi\_neo\_pattern7\_v1.7\_10yrs.db \\
1.000 & 1.002 & 1.7 & twi\_pairs/twi\_pairs\_mixed\_repeat\_v1.7\_10yrs.db \\
1.000 & 1.000 & 1.7 & twi\_pairs/twi\_pairs\_mixed\_v1.7\_10yrs.db \\
0.997 & 0.999 & 1.7 & twi\_pairs/twi\_pairs\_repeat\_v1.7\_10yrs.db \\
1.000 & 1.000 & 1.7 & twi\_pairs/twi\_pairs\_v1.7\_10yrs.db \\
0.992 & 0.993 & 1.7 & u\_long/u\_long\_ms\_30\_v1.7\_10yrs.db \\
0.984 & 0.991 & 1.7 & u\_long/u\_long\_ms\_40\_v1.7\_10yrs.db \\
0.975 & 0.983 & 1.7 & u\_long/u\_long\_ms\_50\_v1.7\_10yrs.db \\
0.967 & 0.978 & 1.7 & u\_long/u\_long\_ms\_60\_v1.7\_10yrs.db \\
1.009 & 1.007 & 1.7 & wfd\_cadence\_drive/cadence\_drive\_gl100\_gcbv1.7\_10yrs.db \\
1.009 & 1.007 & 1.7 & wfd\_cadence\_drive/cadence\_drive\_gl100v1.7\_10yrs.db \\
0.990 & 0.996 & 1.7 & wfd\_cadence\_drive/cadence\_drive\_gl200\_gcbv1.7\_10yrs.db \\
0.982 & 0.991 & 1.7 & wfd\_cadence\_drive/cadence\_drive\_gl200v1.7\_10yrs.db \\
1.014 & 1.011 & 1.7 & wfd\_cadence\_drive/cadence\_drive\_gl30\_gcbv1.7\_10yrs.db \\
1.014 & 1.011 & 1.7 & wfd\_cadence\_drive/cadence\_drive\_gl30v1.7\_10yrs.db \\
1.012 & 1.010 & 1.7.1 & new\_rolling/baseline\_nexp2\_v1.7.1\_10yrs.db \\
0.997 & 1.009 & 1.7.1 & new\_rolling/bulge\_roll\_scale0.90\_nslice2\_fpw0.9\_nrw1.0v1.7\_10yrs.db \\
0.989 & 1.002 & 1.7.1 & new\_rolling/bulge\_roll\_scale0.90\_nslice3\_fpw0.9\_nrw1.0v1.7\_10yrs.db \\
1.013 & 1.024 & 1.7.1 & new\_rolling/footprint\_6\_v1.7.1\_10yrs.db \\
1.019 & 1.026 & 1.7.1 & new\_rolling/full\_disk\_scale0.90\_nslice2\_fpw0.9\_nrw1.0v1.7\_10yrs.db \\
1.011 & 1.020 & 1.7.1 & new\_rolling/full\_disk\_scale0.90\_nslice3\_fpw0.9\_nrw1.0v1.7\_10yrs.db \\
1.036 & 1.040 & 1.7.1 & new\_rolling/full\_disk\_v1.7\_10yrs.db \\
0.981 & 0.983 & 1.7.1 & new\_rolling/rolling\_nm\_scale0.90\_nslice2\_fpw0.9\_nrw1.0v1.7\_10yrs.db \\
0.972 & 0.975 & 1.7.1 & new\_rolling/rolling\_nm\_scale0.90\_nslice3\_fpw0.9\_nrw1.0v1.7\_10yrs.db \\
0.992 & 0.999 & 1.7.1 & new\_rolling/six\_stripe\_scale0.90\_nslice6\_fpw0.9\_nrw0.0v1.7\_10yrs.db \\\enddata
\end{deluxetable*}

\section{Conclusions and Looking Forward \label{sec-conclusions}}

Our simulations indicate that LSST will make a significant contribution to the study of brown dwarfs by measuring thousands of L and T dwarf parallaxes. However, additional spectroscopy and infrared photometry from other sky surveys or targeted follow up will be needed to reliably classify and fully characterize the ultracool dwarfs. This parallax selected sample will enable numerous studies of brown dwarf properties, including a more statistically robust luminosity function. 

Most LSST observing strategy choices only affect the parallax yield at the few percent level. Fundamentally, typical WFD strategies will get $\sim 350$ $z$ and $y$ observations over ten years, and as long as the timing of these visits are not chosen to select against high parallax factors, we can expect to measure parallaxes with roughly mas uncertainties. The simulations that change the sample size (volume) by $\pm10\%$ can be considered unrealistic as they either fail the main SRD requirements for WFD or exclude mini-surveys outside the WFD area.

\begin{figure}
\plotone{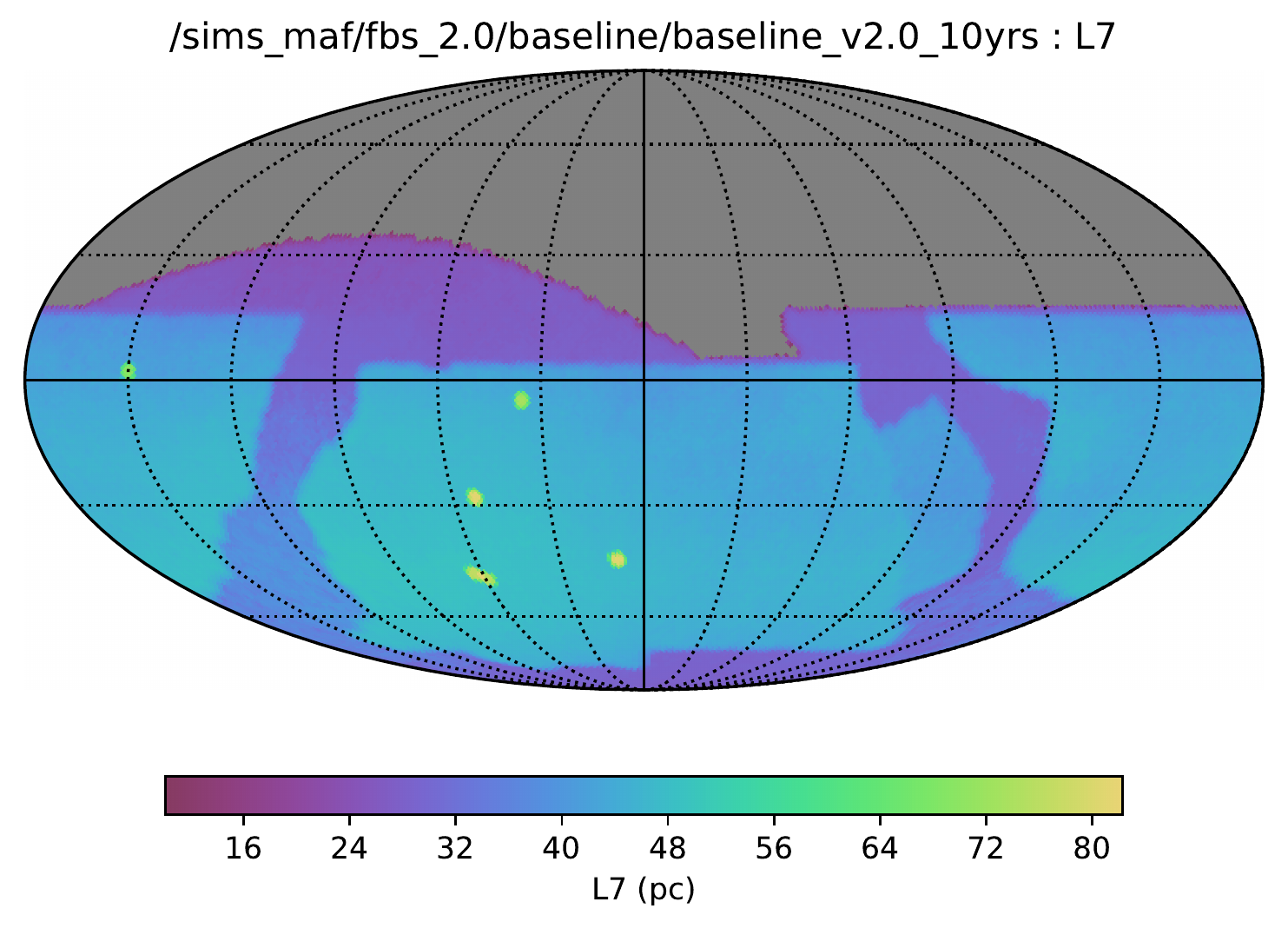}
\caption{\texttt{MAF} sky map of the limiting distance for the new V2.0 baseline.  Note the shift of WFD to low extinction regions, including into the northern hemisphere, and the changed coverage of the Galactic Bulge and Plane. \label{fig-base20-L7-sky} }
\end{figure}

The SCOC has recommended a new baseline (V2.0) which shifts the WFD footprint to minimize interstellar extinction and follows some of the ``big sky" strategy to increase the area of WFD but aiming for the minimal 825 visits.\footnote{The SCOC report is available at \url{https://pstn-053.lsst.io}} The V2.0 baseline results for L7 are illustrated in Figure~\ref{fig-base20-L7-sky}. We find that the increase in areal coverage is outweighed by the loss of visits and other small changes, giving metrics of $F_{\rm L4}=0.95$ and $F_{\rm L7} = 0.97$ relative to the V1.7 baseline; this slight degradation is seen also in the generic parallax metrics. Still, this loss is only a few percent and does not jeopardize the scientific value of the brown dwarf parallax sample, and as noted previously, shifting to 1x30s exposures or improved astrometric uncertainty could improve the sample size. There is real tension with strategies that use the beginning or ends of nights to get additional observations of objects that transit near midnight or use twilight primarily for short-exposure NEO searches, thereby eliminating the WFD observations at high parallax. Additional, more realistic, investigations of LSST astrometry using simulations and commissioning data will be important to understand the full consequences of survey strategy choices and refine the predictions of parallax yield.

\begin{acknowledgments}

We wish to thank Dave Monet for many conversations about parallaxes and his longstanding work advocating for astrometry with LSST. 
The SMWLV working group on survey strategy played a key role in encouraging this work. J.G. thanks Federica Bianco, Gordon Richards, and Weixiang Yu for their help with \texttt{MAF}. 

This paper was created in the nursery of the Vera C. Rubin Legacy Survey of Space Time Science Collaborations and particularly of the Stars, Milky Way, and Local Volume Science Collaboration (SMWLV SC). The authors acknowledge the support of the Vera C. Rubin Legacy Survey of Space and Time SMWLV SC that provided opportunities for collaboration and exchange of ideas and knowledge.

J.G. and D.H. acknowledge support by an LSST Corporation Enabling Science Grant. The authors acknowledge the support of the LSST Corporation that enabled the organization of many workshops and hackathons throughout the cadence optimization process through private fundraising.

Papers in this Focus Issue are supported by the Preparing for Astrophysics with LSST Program, funded by the Heising–Simons Foundation through grant 2021-2975, and administered by Las Cumbres Observatory.

This research uses services or data provided by the Astro Data Lab at NSF's National Optical-Infrared Astronomy Research Laboratory. NOIRLab is operated by the Association of Universities for Research in Astronomy (AURA), Inc. under a cooperative agreement with the National Science Foundation. This research has made use of NASA's Astrophysics Data System Bibliographic Services. 

\end{acknowledgments}

\facilities{Rubin}

\software{Astropy \citep{2013A&A...558A..33A,2018AJ....156..123A},  
          NumPy \citep{2020Natur.585..357H}, 
          Matplotlib \citep{2007CSE.....9...90H},
          Metric Analysis Framework (\texttt{MAF}) \citep{2014SPIE.9149E..0BJ}
          }

\bibliography{lsstp}
\bibliographystyle{aasjournal}

\end{document}